\documentclass[twocolumn]{article}
\usepackage{epsfig}
\begin{document}

\title{\bf\large Observation of the decay $\phi\to\omega \pi^0$ }

\author{M.N.Achasov, S.E.Baru, A.V.Berdyugin, A.V.Bozhenok,\\
D.A.Bukin, S.V.Burdin, T.V.Dimova, S.I.Dolinsky,\\
V.P.Druzhinin \thanks{e-mail: druzhinin@inp.nsk.su},
M.S.Dubrovin, I.A.Gaponenko, V.B.Golubev,\\
V.N.Ivanchenko, A.A.Korol, M.S.Korostelev, S.V.Koshuba,\\
E.V.Pakhtusova, A.A.Polunin, E.E.Pyata, A.A.Salnikov,\\
V.V.Shary, S.I.Serednyakov, Yu.M.Shatunov, V.A.Sidorov,\\
A.N.Skrinsky, Z.K.Silagadze, Yu.S.Velikzhanin
\vspace*{3mm} \\ Budker Institute of Nuclear Physics,  630 090,
Novosibirsk, Russia, \\ \vspace*{1mm}
Novosibirsk State University, 630 090, Novosibirsk, Russia  }

\date{}
\maketitle

\begin{abstract}
  The reaction  $e^+e^-\to\omega\pi^0\to\pi^+\pi^-\pi^0\pi^0$ 
has been studied
with SND detector at VEPP-2M  $e^+e^-$ collider in the vicinity of 
the $\phi$ meson
resonance. The observed interference pattern in the energy dependence 
of the cross section is consistent with
existence of the decay $\phi\to\omega \pi^0$ with a branching ratio of
$$B(\phi\to\omega \pi^0)=(4.8^{+1.9}_{-1.7}\pm0.8)\cdot 10^{-5}.$$
The real and imaginary parts of the decay amplitude were measured. The 
$\phi\to\omega \pi^0$ decay was observed for the first time.
\end{abstract}


\section{Introduction}
Recently in experiments with SND and CMD-2 detectors at VEPP-2M
$e^+e^-$ collider the study
of $\phi$-meson rare decays with branching ratios of 10$^{-4}$--10$^{-5}$
became available \cite{Serednyakov/H-97,Prep.97,CMD-2}. One of
such  decays is an OZI and G-parity violating $\phi\to\omega\pi^0$
decay. The predicted branching fraction of this decay is of the order of
$5\cdot 10^{-5}$ \cite{theor1,theor2}. It varies within wide limits
depending on the nature of  $\rho$, $\omega$ and $\phi$-meson
mixing and existence of
direct  $\phi\to\omega\pi^0$ transition. Since the
nonresonant cross section  of the process $e^+e^-\to\omega\pi^0$
in the vicinity of the $\phi$-meson peak is relatively large, 
the decay $\phi\to\omega\pi^0$ reveals itself as an  interference pattern
in the cross section energy dependence. This allows to extract from the data 
both real and imaginary parts of the decay amplitude.

    The  $\phi\to\omega\pi^0$ decay has not been observed yet.
In our preliminary study \cite{Serednyakov/H-97,Prep.97}
an upper limit  for the decay probability was placed
at a level of the theoretical prediction of $5\cdot 10^{-5}$.

\section{Experiment}
   The experiment was carried out with SND detector at VEPP-2M 
 in 1996--1997. SND  is a general purpose non-magnetic
detector \cite{SND}, based on 3.6~t three layer spherical electromagnetic 
calorimeter with 1620 NaI(Tl) crystals covering 90\% of $4\pi$ solid angle.
The energy resolution for electromagnetic showers is
$\sigma_E/E=4.2\%/\sqrt[4]{E(GeV)}$, the angular resolution is about
1.5$^\circ$.  The angles of charged particles are measured by two cylindrical
drift chambers covering 95\% of  $4\pi$ solid angle. The
angular accuracy of charged track measurements is about 0.4$^\circ$ and
2.0$^\circ$ in azimuth and polar directions respectively.

Seven successive data taking runs were performed at 14
energy points in the energy range 2E=980--1060 MeV. The total
integrated luminosity  $\Delta$L=3.7 $pb^{-1}$ corresponds
to $6.3\cdot 10^{6}$ produced  $\phi$ mesons. The
integrated luminosity was measured using $e^+e^-\to e^+e^-$ and
$e^+e^-\to\gamma\gamma$ reactions with the accuracy of 3\%.
  
\section{Event selection}
For a search for the $\phi\to\omega\pi^0$ decay
we studied the cross section of the process
\begin{equation}
e^{+}e^{-} \to\omega\pi^0\to\pi^+\pi^-\pi^0\pi^0
\label{ompi}
\end{equation}
in the vicinity of the  $\phi$-meson. Events with
two charged particles and four or more  photons were selected for analysis.
To suppress beam background 
the production point of charged particles was required to be 
within 0.5 cm from the detector center
in the  azimuth plane and $\pm$7.5 cm along the beam direction
(the longitudinal size of the interaction region  $\sigma_{z}$ is about 2 cm).

 Since for the process under study
a probability  to find  spurious photons in  
the events is rather large (about 15\%),
nearly all main $\phi$-meson decays are a source of background
in our search:
\begin{equation}
 e^{+}e^{-} \to\phi\to K^{+}K^{-};
\label{kkc}
\end{equation}
\begin{equation}
 e^{+}e^{-} \to\phi\to K_{S}K_{L}, K_{S} \to\pi^{+}\pi^{-};
\label{kkn}
\end{equation}
\begin{equation}
 e^{+}e^{-} \to\phi\to\pi^{+}\pi^{-}\pi^0;
\label{3pi}
\end{equation}
\begin{equation}
 e^{+}e^{-} \to\phi\to\eta\gamma, \eta\to\pi^{+}\pi^{-}\pi^0.
\label{etag}
\end{equation}
      The main source of the nonresonant background is the  process
\begin{equation}
 e^{+}e^{-} \to\pi^+\pi^-\pi^0\pi^0
\label{4pi}
\end{equation}
which has the same final state as the process (\ref{ompi}), but 
without the  $\omega\pi^0$ intermediate state. The interference between
the processes (\ref{ompi}) and (\ref{4pi}) is significant only in the small
part of the decay phase space, where $\pi^+\pi^-\pi^0$ invariant mass is
within the $\omega$-meson width. Estimated contribution of the interference
term in the energy region under study does not exceed 2\%
of the cross section of the process (\ref{ompi}) and was neglected in the 
further analysis. It was treated instead as an additional source of 
systematic error in the nonresonant cross section of the process (\ref{ompi}).

  To suppress the background from the processes  (\ref{kkc}) and (\ref{kkn}),
the following cuts were imposed: 
\begin{itemize}
\item average $dE/dx$ losses in the drift chamber
$dE/dx<4\cdot (dE/dx)_{min}$, where
$(dE/dx)_{min}$ are average $dE/dx$ losses of a minimum
ionizing particle;
\item the spatial angle between charged particles is less than $140^{\circ}$.
\end{itemize}
The first condition suppresses events of the process (\ref{kkc})
with slow ($\beta\approx 0.25$) charged kaons having
large $dE/dx$ losses in the SND drift chamber.  The second condition suppresses
events from the process (\ref{kkn})
with a minimum angle between pions from the  $K_S$ decay close to 150 degrees.

For each selected event independent kinematic fits were performed 
under three following assumptions about the reaction mechanism:
\begin{enumerate}
\item The event originates from the process  $e^+e^-\to\pi^+\pi^-\pi^0$.
The value of the likelihood function $\chi_{3\pi}$ is calculated.
\item The event originates from the process  $e^+e^-
\to\pi^+\pi^-\pi^0\gamma$. The photon recoil mass $M_{rec}$
is calculated.
\item The event is due to the process $e^+e^-\to\pi^+\pi^-\pi^0\pi^0$.
The value of the likelihood function $\chi_{4\pi}$ is calculated together with
$M_{3\pi}$ -- the recoil mass of  the $\pi^0$-meson 
closest to the $\omega$-meson mass.
\end{enumerate}
    In the events where the number of photons  exceeds that required by
a certain hypothesis, extra photons are considered as 
spurious ones and rejected.
To do that all possible subsets with a correct number of
photons were tested and the one, corresponding to a maximum likelihood
of a certain hypothesis, was selected.
    
\begin{figure}
\epsfig{figure=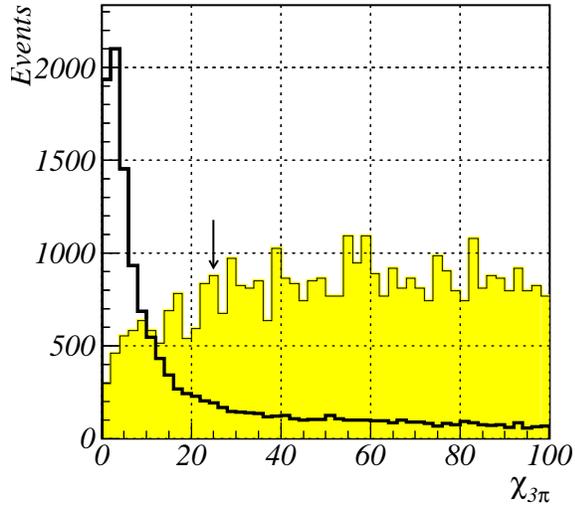}
\caption {Distribution over $\chi _{3\pi}$ for experimental
events and simulation of the process 
$e^{+}e^{-} \to\omega\pi^0\to\pi^+\pi^- 2\pi^0$ (shaded histogram). 
The cut is indicated by the arrow.}
\label{f1}
\end{figure}

\begin{figure}
\epsfig{figure=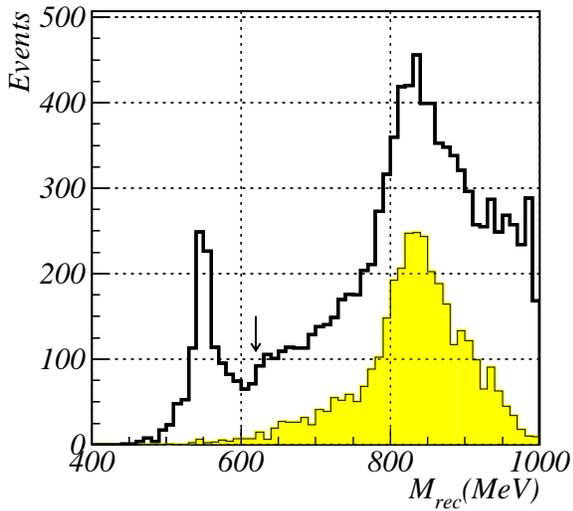}
\caption { Distribution over the parameter $M_{rec}$ for experimental
events and simulation of the process
$ e^{+}e^{-} \to\omega\pi^0\to\pi^+\pi^- 2\pi^0$ (shaded histogram). 
The cut is indicated by the arrow.}
\label{f2}
\end{figure}

    The distribution of experimental and simulated events of the process 
(\ref{ompi}) over the parameter $\chi_{3\pi}$, is shown in Fig. \ref{f1}.
One can see a considerable contribution from the process (\ref{3pi}),
producing a peak at low $\chi_{3\pi}$. Fig. \ref{f2}
shows the experimental $M_{rec}$ spectrum, where a clear $\eta$-meson peak
from the reaction (\ref{etag}) is seen.
To suppress the background from the processes (\ref{3pi}) and
(\ref{etag}) the following cuts were applied:
$$\chi_{3\pi}>25,\,\,M_{rec}>620\,MeV.$$

\section{Data analysis}

\begin{figure}
\epsfig{figure=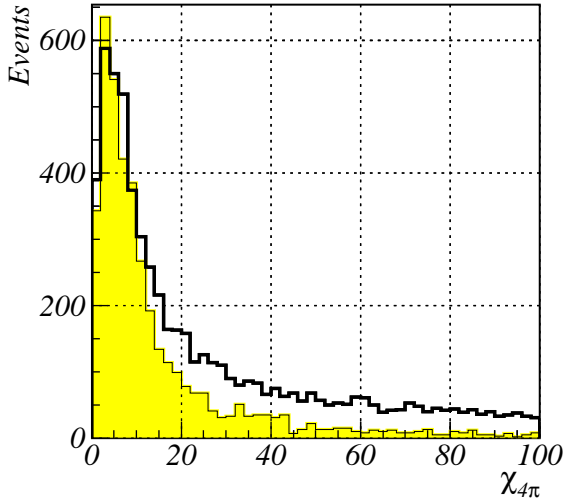}
\caption{ Distribution over $\chi_{4\pi}$ for experimental events
and simulation of the process  (\ref{ompi}) (shaded histogram).}
\label{f3}
\end{figure}

\begin{figure}
\epsfig{figure=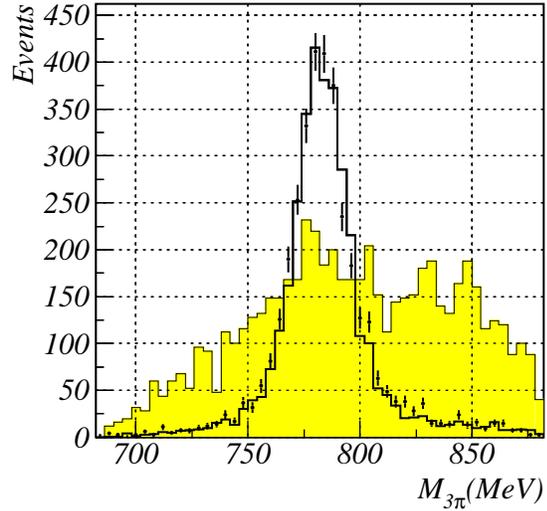}
\caption{ Distribution over parameter $M_{3\pi}$ for experimental events
(points with errors), simulation of the process (\ref{4pi}) 
(shaded histogram), and
simulation of the process (\ref{ompi}) with  an 11\% 
admixture of the process (\ref{4pi}) (histogram).}
\label{f4}
\end{figure}

Figure \ref{f3} shows the distribution of
experimental and simulated events of the process (\ref{ompi}) over
$\chi_{4\pi}$. Considerable difference between the tails of measured 
and simulated spectra indicates the presence of 
a background surviving the cuts. One can see that the experimental peak in
Fig. \ref{f3} is broader than a simulated one. It means that the
simulation of the $\chi_{4\pi}$ distribution is not precise.
In Fig.\ref{f4} the distributions over the  $\pi^0$ 
recoil mass $M_{3\pi}$ for experimental events with $\chi_{4\pi}<20$, 
simulated events of the process (\ref{4pi}) (the $\rho\pi\pi$ mechanism
was assumed), and simulation of the process  (\ref{ompi}) with the 11\% 
admixture of the process  (\ref{4pi}) are presented. 
The last distribution is in good
agreement with the experimental one.
For further analysis two additional cuts were applied: 
$$
\chi_{4\pi}<40,\,|M_{3\pi}-782|<100.
$$
The total of 4500 events which  survived the cuts were divided into 4 classes:
\begin{enumerate}
\item $\chi_{4\pi}<20,\, |M_{3\pi}-782|<35$;
\item $\chi_{4\pi}<20,\, |M_{3\pi}-782|>35$;
\item $\chi_{4\pi}>20,\, |M_{3\pi}-782|<35$;
\item $\chi_{4\pi}>20,\, |M_{3\pi}-782|>35$;
\end{enumerate}
with the relative populations of 0.69:0.09:0.16:0.06 respectively.

   The visible cross section for each class $\sigma_{\mathrm{vis}\,\mathit{i}}$
was fitted according to the following formulae:
\begin{eqnarray}
\label{eq1}
&\sigma_{\mathrm{vis}\,\mathit{i}}(E) = \alpha_i\sigma_{\omega\pi}(E)+
\beta_i\sigma_{4\pi}(E)+\sigma_{\phi i}(E),&\nonumber \\
&\sigma_{\omega\pi}(E) = \varepsilon B(\omega\to 3\pi) 
(\sigma_0+A(E-m_\phi))\times &\nonumber\\
&\biggl |1-Z \frac{m_\phi\Gamma_\phi}{D_\phi}\biggr |^2(1+\delta).&
\end{eqnarray}
Here $\sigma_{\omega\pi}$, $\sigma_{4\pi}$ are the visible cross sections
of the processes
(\ref{ompi}) and (\ref{4pi}), $\alpha_i$, $\beta_i$ are
the probabilities for events of the processes (\ref{ompi})
or (\ref{4pi}) to be found in the $i$-th class 
($\sum\alpha_i=1$ and $\sum\beta_i=1$
at $E=m_\phi$), $\sigma_{\phi i}$ is the visible cross section of the 
resonant background in the $i$-th class, $\sigma_0$ is the nonresonant 
cross section of the process $e^{+}e^{-}\to\omega\pi^0$ at $E=m_\phi$, 
$A$ is its slope, $\varepsilon$ is the detection efficiency for the process 
(\ref{ompi}) at $E=m_\phi$, $Z$ is the complex interference amplitude,
$m_\phi,\Gamma_\phi$, $D_\phi=m_\phi^2-E^2-iE\Gamma_\phi(E)$ are the
mass, width and $\phi$-meson propagator function respectively, 
$B(\omega\to 3\pi)=0.888$ is
the branching ratio of the $\omega\to 3\pi$ decay\cite{PDG},
$\delta$ is a radiative correction calculated according to \cite{RadCor}.

  The fitting was performed for all 4 classes simultaneously.
The class 1 with a small resonant background was the most important for the 
determination of $\sigma_0$, $A$, $Z$. Classes 2--4 were used to determine
the background from $\phi$ decays: $\sigma_{\phi 2}$, $\sigma_{\phi 3}$,
$\sigma_{\phi 4}$. In the fit it was assumed  that for the resonant background
the distribution over  $M_{3\pi}$ is independent of $\chi_{4\pi}$,
and therefore the background for the class 1 can be
obtained from the expression:
$\sigma_{\phi 1}=\sigma_{\phi 2}\cdot(\sigma_{\phi 3}/\sigma_{\phi 4})$.
The validity of this relation was checked with the statistical
accuracy of 15\%
for simulated events and specially selected experimental 
events  of the processes (\ref{kkc})--(\ref{etag}). The cross section of 
the resonant background $\sigma_{\phi 1}$ in the resonance maximum was 
found to be $\sigma_{\phi 1}=(31\pm 17)~pb$, that is  about 4\% 
of the visible cross section $\sigma_{\mathrm{vis}\,\mathit{1}}$ in the class 1.

  The contribution of the process  (\ref{4pi}) was determined using the 
value of
the nonresonant part of the cross section in the class 2 and the ratios
$\alpha_2/\alpha_1=0.058$ and $\beta_2/\beta_1=1.05$, obtained from
simulation.  The cross section  $\sigma_{4\pi}(E)$ was fitted by a sum
of a linear function and an interference term similar to
$\sigma_{\omega\pi}(E)$ in the expression (\ref{eq1}).
For the  OZI-rule and G-parity double suppressed
$\phi\to\pi^+\pi^-\pi^0\pi^0$ decay, the main mechanism is a transition of
the $\phi$ meson into $\pi^+\pi^-\pi^0\pi^0$ via a virtual photon. 
The calculated real part of the interference amplitude  $Z$ 
due to this decay mechanism is equal to 0.127. 
In the fit the real and imaginary parts of the interference amplitude of the
process (6) were set to 0.127 and 0.0 respectively, while their errors were
set to 0.07 and 0.15 in agreement with the accuracy of theoretical estimations
and experimental data on another double suppressed decay $\phi\to\pi^+\pi^-$.
 The nonresonant cross section
(\ref{4pi}) was found to be equal to $(11\pm4)$\% 
of the cross section of the process (\ref{ompi}).
In the class 1 the contribution of this process is smaller: $4.7\%$.

  Since the simulated distribution over the $\chi_{4\pi}$ may differ from
the experimental one,  the coefficients  $\alpha_1$ and  $\alpha_3$ were
determined from the fit.
Other coefficients  $\alpha_i$ and  $\beta_i$ were derived from the
relations described above, with the  
additional assumption  that the distributions over $\chi_{4\pi}$ in
the processes  (\ref{ompi}) and (\ref{4pi}) are the same.
The coefficients  $\alpha_i$, $\beta_i$ vary slowly with energy, 
and their energy dependence approximated by linear functions
was obtained by simulation.

 The detection efficiency  $\varepsilon$ for the process (\ref{ompi}) was
also obtained from simulation. To estimate a systematic error in $\varepsilon$,
we processed data with softer cuts and studied the class of events
with one charged particle and four photons. It was found, that the
detection efficiency obtained by Monte Carlo simulation
must be corrected by -11\%. 
The correction is mainly due to
the inaccuracy in the simulation 
of $\chi_{4\pi}$ distribution and loss of charged
particles during track reconstruction. With this correction, the detection
efficiency is equal to  $17.6\pm1.8\%$ with a systematic error of 10\%. 

    The total number of fit parameters describing energy dependence of the
cross sections in all 4 classes of selected events is 11. For each class
the cross section was measured in 14 energy points. At
$\chi^2/n_D=43/45$ the following values of main fit
parameters were obtained:
\begin{eqnarray}
\label{res1}
&\sigma_0=(8.2\pm 0.2\pm 0.9) nb,&\nonumber \\
&A=(0.088\pm 0.009\pm 0.011) nb/MeV,&\\
&Re(Z)=0.104\pm0.028\pm 0.006,&\nonumber\\
&Im(Z)=-0.118\pm0.030\pm 0.009,&\nonumber
\end{eqnarray}
where the first error is a statistical one obtained during the
fitting, and the second is systematic to be discussed below.

\begin{figure}
\epsfig{figure=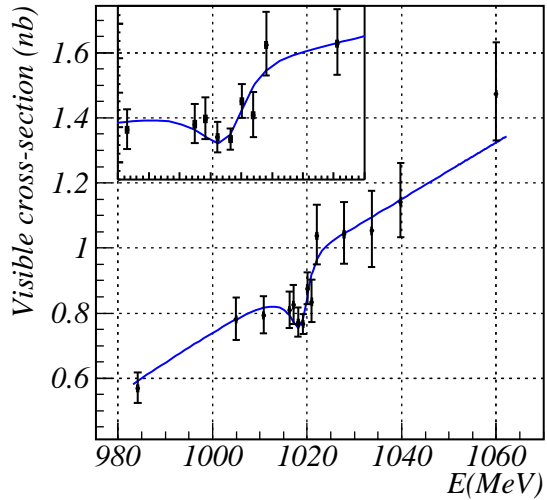}
\caption{Energy dependence of the visible cross section for the process
(\ref{ompi}) and optimal curve describing data.}
\label{f5}
\end{figure}

 The visible cross section for the class 1 and the fitted curve are shown in
Fig. \ref{f5}. Despite a 4\% 
resonant background, the
interference pattern is clearly seen. Another representation of the
interference amplitude is $Z=|Z|\cdot e^{\psi}$ with
\begin{eqnarray}
\label{res2}
&|Z|=0.158\pm0.030\pm0.009,&\\
&\psi=(-49\pm10\pm4)^{\circ}.&\nonumber
\end{eqnarray}

     The branching ratio for the decay  $\phi\to\omega\pi^0$ can be
obtained from the following relation:
$$B(\phi\to\omega\pi^0)=\frac{\sigma_0\cdot|Z|^2}{\sigma_\phi}
=(4.8^{+1.9}_{-1.7}\pm0.8)\cdot 10^{-5},$$
where $\sigma_{\phi} =12\pi B(\phi\to e^+e^-)/m_{\phi}^2=4220$ nb is
the cross section in the  $\phi$-meson peak  \cite{PDG}.

The following sources  of systematic errors were considered:
inaccuracy of the detection efficiency estimation, interference
between the processes (\ref{ompi}) and (\ref{4pi}), contribution from the
nonresonant $e^+e-\to 3\pi\gamma$ reaction, inaccuracy of 
Monte Carlo simulation of the $M_{3\pi}$ distribution, and possible deviations
from linear energy dependence of the nonresonant $e^+e-\to 3\pi\gamma$
cross section. The total systematic error in the nonresonant cross section
$\sigma_0$ was estimated to be 11\%.
It is determined mainly by the 10\%
uncertainty in the detection efficiency.

Of the factors listed above, the main systematic error in the measured
interference amplitude $Z$ is introduced by a possible nonlinearity of the
energy dependence of the cross section of the process (\ref{ompi}).
It is equal to 4\%
for   $Re(Z)$ and 7\%
for $Im(Z)$. Other factors contribute to  $Re(Z)$ only at a level of 2\%.
This nonlinearity also determines the error in the slope $A$.

  To estimate systematic errors caused by a possible detector instability
during lengthy data taking, we divided data into three subsets and
processed  them separately. It was found that all three subsets
are consistent with the fit obtained from the summary data.
The fit parameters obtained from each subset agree well with 
each other and with
(\ref{res1}). To check the stability of the obtained results,
we loosened cuts: for the parameter $\chi_{4\pi}$ the cuts
50 and 100 were used
instead of 20 and 40, for the parameter $|M_{3\pi}-782|$ --
 50 and 100 instead of 35 and 100. As a result, the detection
efficiency increased up to 20\%
and the resonant background in the
maximum became $\sigma_{\phi 1}=(0.20\pm0.05)$ nb, that is 20\%
of the visible cross section (\ref{ompi}). However the obtained interference
amplitude was $Re(Z)=0.108\pm0.026$, $Im(Z)=-0.121\pm0.029$
which does not contradict (\ref{res1}), thus confirming the validity of the
background subtraction procedure.
As a final result we take (\ref{res1}), because it was obtained with 
a lower resonant background. The difference between the results was
considered as an estimate of the systematic error of the resonant background
subtraction. Thus, the total systematic error of the interference amplitude $Z$
is equal to 6\%
for $Re(Z)$ and 8\%
for $Im(Z)$.

\section{Conclusion}
 The nonresonant cross section
of the process $e^+e^-\to\omega\pi^0$ obtained in this work
 is in agreement with our old result
\cite{Ph.Rep} on the neutral $\omega$-meson decay mode:
$\sigma(e^+e^-\to\omega\pi^0\to\pi^0\pi^0\gamma)$=
$(8.7\pm1.0\pm0.7)$~nb. The measured nonresonant cross section  $\sigma_0$
is almost twice higher than the expected value, calculated with only
$\rho(770)\to\omega\pi^0$ transition taken into account.
The agreement may be significantly improved if the known radial excitations of
the $\rho$ meson are taken into account.

The measured interference amplitude  $Z$ (\ref{res1})
is close to the lower limit of theoretical predictions \cite{theor2},
although contributions from radial excitations of $\rho$
not considered in \cite{theor2}, may affect its value. 
Another important remark is a small value of the measured
real part of the interference amplitude $Re(Z)$, which could be hardly
explained by the standard $\phi-\omega$ mixing model \cite{theor2}.
For example, the value of $8.2\cdot10^{-5}$
 predicted in \cite{theor1} for the branching ratio of
the decay $\phi\to\omega\pi^0$,  is
1.5 times higher than one measured in this work.

  The interference amplitude  (\ref{res2})  measured in this work is five
standard deviations above zero. Thus, we claim the existence of the
decay $\phi\to\omega\pi^0$ with the branching ratio of
$$B(\phi\to\omega \pi^0)=(4.8^{+1.9}_{-1.7}\pm0.8)\cdot 10^{-5}.$$

\section{Acknowledgement}
The authors are grateful to S.I.Eidelman for useful discussions and 
valuable comments. 
This work is supported in part by The Russian Fund for Basic 
Researches, grants No. 97-02-18561 and 96-15-96327.

\clearpage

\end{document}